\documentclass[twoside]{article}

\usepackage{amssymb}
\usepackage{amsmath}
\usepackage{graphicx}
\usepackage{hyperref}
\usepackage{qic}

\begin{document}
\setlength{\textheight}{8.0truein}    

\runninghead{Integer arithmetic with hybrid quantum-classical circuits}
            {C. M. Maynard and E. Pius}

\normalsize\textlineskip
\thispagestyle{empty}
\setcounter{page}{1}


\vspace*{0.88truein}

\alphfootnote

\fpage{1}

\centerline{\bf INTEGER ARITHMETIC}
\vspace*{0.035truein}
\centerline{\bf WITH HYBRID QUANTUM-CLASSICAL CIRCUITS}

\vspace*{0.37truein}

\centerline{\footnotesize CHRISTOPHER M. MAYNARD${}^{a,b}$}
\centerline{\footnotesize\it ${}^a$The Met Office, FitzRoy Road}
\baselineskip=10pt
\centerline{\footnotesize\it Exeter, EX1 3PB, United Kingdom}
\vspace*{10pt}

\centerline{\footnotesize EINAR PIUS${}^b$}
\vspace*{0.015truein}
\centerline{\footnotesize\it ${}^b$Edinburgh Parallel Computing Centre, School of Physics, The University of Edinburgh}
\baselineskip=10pt
\centerline{\footnotesize\it Edinburgh, EH9 3JZ, United Kingdom}
\vspace*{0.225truein}

\vspace*{0.21truein}

\abstracts{
	Quantum circuits which perform integer arithmetic could potentially
	outperform their classical counterparts. In this paper, a
	quantum circuit is considered which performs a specific
	computational pattern on classically represented integers to
	accelerate the computation.  Such a hybrid circuit could be embedded
	in a conventional computer architecture as a quantum device or
	accelerator.  In particular, a quantum multiply-add circuit (QMAC)
	using a Quantum Fourier Transform (QFT) is proposed which can
	perform the calculation on conventional integers faster than its
	conventional counterpart.
	Whereas classically applying a multiply-adder (MAC) $n$ times
	to $k$ bit integers would require $O(n \log k)$ parallel steps,
	the hybrid QMAC needs only $O(n + k)$ steps for the exact result and
	$O(n + \log k)$ steps for an approximate result. 
}{}{}

\vspace*{10pt}

\vspace*{3pt}
\communicate{to be filled by the Editorial}

\vspace*{1pt}\textlineskip    

\section{Introduction}
Quantum computing has the potential to dramatically change the nature
of computing, but has mostly been a theoretical subject partly due to
the difficulties in building physical quantum circuits. However,
recent progress has enabled the first, albeit small, quantum devices
to be constructed, see for example~\cite{Bonneau:2012} utilising
photonics. These devices are not complete quantum computers, but
consist of simple quantum circuits capable of processing information to solve
specific problems. Critically, these devices can be fabricated in
silicon which could lead to their integration with conventional
microelectronics. How would such a hybrid of conventional and quantum
microprocessor be used? Co-processor architectures have been developed
in the past but perhaps the most promising context would be to
consider the quantum device as an accelerator.

There are several examples of modern heterogeneous computer
architectures. For example, Graphical Processing Units (GPUs) have
been used extensively in the field of scientific numerical computing
to accelerate specific aspects of these calculations, where some
suitably defined compute kernel is offloaded from the CPU and executed
faster on the GPU.  Another analogy can be drawn with field
programmable gate arrays (FPGAs) where particular computational
patterns in software can be instantiated in hardware using the
reprogrammable logic of these devices, see for
example~\cite{Baxter:2007, Almer:2009}. Rather than accelerating an
entire kernel as would be required for a GPU, a quantum device could
be employed to accelerate a specific computational pattern. Moreover,
as this device would function as an accelerator, a complete quantum
computer would not be required.  Furthermore, the effects of quantum
decoherence which destroys quantum information can be mitigated
because such quantum circuits need only to be in an entangled state
for brief period.

The addition and multiplication of small integers are the simplest
computational patterns. Here, the manipulation of $n$ classically
represented integers of size $k$ bits by a quantum circuit is
considered.  A key consideration is the number of parallel steps it
takes to execute a quantum circuit, {\em i.e.} the depth of the
quantum circuit implementing the computation.  The first quantum
addition circuit was proposed by Vedral {\em et. al.} in 1995
\cite{Vedral:1996}. It is a quantum version of the classical ripple
carry adder\footnote{ see for example \cite{Smith:1997} for a textbook
  on circuit design}.  The quantum ripple carry adder has been further
studied in the circuit model of quantum
computing~\cite{Cheng:2002b,Cuccaro:2004,Chakrabarti:2008,Tagahashi:2010}
and in the Measurement Based Quantum Computing Model
(MBQC)~\cite{Raussendorf:2003}. The quantum circuits implementing the
classical carry-lookahead
\addtocounter{footnote}{-1}adder\footnotemark have been investigated
in \cite{Draper:2006,Chakrabarti:2008,Jamal:2012} and the MBQC design
in \cite{Trisetyarso:2009b}.

Most of the quantum adders constructed are thus quantum versions of
classical ripple carry or carry-lookahead adders. A notable exception
is the addition circuit proposed by Draper in \cite{Draper:2000},
which utilises the quantum Fourier transform (QFT) operation. Whilst this
particular circuit performs no better than a classical carry-lookahead
circuit, employing circuit features which are specific to quantum circuits
rather than quantum analogues of classical circuits may allow performance
gains to be achieved. 

Quantum arithmetic circuits for integer multiplication have been
proposed in \cite{AlvarezSanchez:2008, Thapliyal:2006a}, but this is
the first work studying the multiply-add operation in the quantum
setting.  Although quantum arithmetic logic units (ALUs) have been
proposed in several papers \cite{Thapliyal:2005, Fahdil:2010,
  Thomsen:2010, Syamala:2011}, none of them analyse if the addition
and multiplication could be merged into a single, more efficient
multiply-add operation. 

In this work, the QFT, highly entangled quantum states obtained
through "fanning-out"~\cite{Moore:2001} of the QFT states, and the
classical properties of a hybrid circuit are combined together to
produce a QFT multiply-add circuit (QMAC) for classical integers which
outperforms a conventional multiply-add unit.

The rest of the paper is organised as follows: In Section~\ref{sec:QMAC}
the QMAC is described. In Section~\ref{sec:Analysis} the depth of the
circuit is analysed and compared to a conventional multiply-add circuit.
Finally, in Section~\ref{sec:Results} the results are presented.

\section{The QFT Multiply-Add Circuit}
\label{sec:QMAC}

\noindent
Consider a unitary operator, $M$, which when combined with the QFT
can be used to compute the action of a classical integer MAC: $z + y \cdot x$, where $z,y,x \in \mathbb{Z}$.
This operator is then decomposed into single qubit gates.
The decomposition shown is particularly useful, since it allows for the construction
of a parallel hybrid circuit which is presented at the end of this section.
For the sake of notational simplicity, only unsigned integers are considered
but the presented circuits can work with signed integers if the two's complement representation is used.

Let $z_k \cdots z_2 z_1$ be the binary representation of a $k$ bit
integer $z$ such that $z = z_k 2^{k-1} + \cdots + z_2 2^1 + z_1 2^0$ and $0.z_k \cdots z_2 z_1$ the binary fraction $z_k/2^1
+ \cdots + z_{2} / 2^{k-1} + z_1 / 2^{k}$. Then the $QFT$ of a $k$ qubit computational basis state $| z
\rangle$ can be written as \cite{Nielsen:2011}:
\begin{align}
	QFT | z \rangle = &\frac{1}{\sqrt{2}}(|0 \rangle + e^{2 \pi i 0.z_1}|1 \rangle) \nonumber \\
	\otimes &\frac{1}{\sqrt{2}}(|0 \rangle + e^{2 \pi i 0.z_2 z_1}|1 \rangle) \nonumber \\
	\otimes &\cdots \nonumber \\
	\otimes &\frac{1}{\sqrt{2}}(|0 \rangle + e^{2 \pi i 0.z_k \cdots z_2 z_1}|1 \rangle).
\end{align}
Let $M_j(y, x)$ be a single qubit unitary operator defined as follows: 
\begin{align}
	M_j(y, x) | 0 \rangle &\rightarrow | 0 \rangle \\
	M_j(y, x) | 1 \rangle &\rightarrow e^{2 i \pi 0.y_j \cdots y_1 \cdot x} | 1 \rangle,
\end{align}
where $x, y \in \mathbb{Z}$ are $k$ bit integers. The effect of applying $M_j(y,x)$ to a state which has
the first $j$ bits of a $k$-bit integer $z$ encoded in its relative phase is
\begin{align}
	M_j(y, x) \frac{1}{\sqrt{2}}(|0 \rangle + e^{2 \pi i 0.z_j \cdots z_2 z_1} | 1 \rangle) = \frac{1}{\sqrt{2}}(|0 \rangle + e^{2 \pi i (0.z_j \cdots z_2 z_1 + 0.y_j \cdots y_2 y_1 \cdot x)} | 1 \rangle) 
\end{align}
The above equation shows that the action of $M_j(y,x)$ is similar to applying a MAC operator to the binary fraction encoded in the relative phase, \emph{i.e.} it multiplies the binary fraction $0.y_j \cdots y_2 y_1$ with $x$ and adds it to  $0.z_j \cdots z_2 z_1$.
The $k$ qubit quantum operator $M$ corresponding to a MAC is defined as
\begin{align}
	\label{eq:qmac}
	M(y, x) = M_1(y, x) \otimes M_2(y, x) \otimes \cdots \otimes M_k(y, x).
\end{align}
The application of $M(y, x)$ to $QFT | z \rangle$ will result in the state
\begin{align}
	M(y, x) QFT | z \rangle =
		&\frac{1}{\sqrt{2}}(|0 \rangle + e^{2 \pi i (0.z_1 + 0.y_1 \cdot x)}|1 \rangle) \nonumber \\
		\otimes &\frac{1}{\sqrt{2}}(|0 \rangle + e^{2 \pi i (0.z_1 z_2 + 0.y_2 y_1 \cdot x)}|1 \rangle) \nonumber \\
		\otimes &\cdots \nonumber \\
		\otimes &\frac{1}{\sqrt{2}}(|0 \rangle + e^{2 \pi i (0.z_k \cdots z_2 z_1 + 0.y_k \cdots y_2 y_1 \cdot x)}|1 \rangle)
		&= QFT | z + y \cdot x \rangle.
\end{align}
Applying the $QFT^\dagger$ operator and measuring the result in the computational basis gives the output $z + y \cdot x$, which would also be the effect of a classical MAC applied to $x$, $y$, $z$. Note that since $e^{2 \pi i (m + 0.z_l \cdots z_2 z_1)} = e^{2 \pi i 0.z_l \cdots z_2 z_1}$ for every $m \in \mathbb{Z}$, $z \in \{ 0, 1 \} ^k$, and $l \in \{ 1, 2, \cdots, k \}$ the output is computed modulo $k$.

Any realistic quantum device would have to be built using quantum
gates acting on a limited number of qubits, thus the $M(y, x)$
operator needs to be decomposed into one- and two-qubit quantum gates.  To
obtain a performance that surpasses classical MACs 
the $M(y, x)$ operation will be constructed in a way that allows every gate
in its circuit to be applied in one simultaneous step. The following gates are used
in the circuit construction:

\begin{align}
	R_j =
	\begin{bmatrix}
		1 & 0 \\
		0 & e^{\frac{2 i \pi}{2^j}}
	\end{bmatrix}, \quad
	CNOT =
	\begin{bmatrix}
		1 & 0 & 0 & 0 \\
		0 & 1 & 0 & 0 \\
		0 & 0 & 0 & 1 \\
		0 & 0 & 1 & 0
	\end{bmatrix},
\end{align}
where $R_j$ is a phase shift gate and $CNOT$ is the
controlled NOT gate. Note that the operator $R_j$ has the following properties:
\begin{align}
	\label{eq:r_identity}
	\forall j \in \mathbb{Z} < 1 \quad R_j &= I \\
	\label{eq:r_pow}
	R_j^{2^m} =
	\begin{bmatrix}
		1 & 0 \\
		0 & e^{\frac{2 i \pi 2^m}{2^j}}
	\end{bmatrix} =
	\begin{bmatrix}
		1 & 0 \\
		0 & e^{\frac{2 i \pi}{2^{j - m}}}
	\end{bmatrix} &=
	R_{j - m}.
\end{align}
The $j$-qubit \emph{fan-out} operator $F_j$ which maps $| a \rangle
| b_1 \rangle \cdots | b_{j-1} \rangle \rightarrow | a \rangle | b_1
\oplus a \rangle \cdots | b_{j-1} \oplus a \rangle$, where $b_i \oplus
a = (b_i \oplus a) \mod 2$ is also required. It is trivial to see that $F^\dagger = F$.  The operator $Q_j(y) = R_{j
\cdot y_1} \cdots R_{2 \cdot y_{j-1}} R_{1 \cdot y_j}$ is used as a sub-circuit
in the $M(y, x)$ construction.  The effect of $Q_j(y)$ on the one
qubit computational basis is:
\begin{align}
	| 0 \rangle &\rightarrow | 0 \rangle \\
	| 1 \rangle &\rightarrow e^{2 \pi i \cdot 0.y_j \cdots y_2 y_1} | 1 \rangle.
\end{align}
Note that since $y_m$, where $m \in \{ 1, 2, \cdots, j \}$, is a binary value and
$R_0 = R^0_l = I$ for every $l \in \mathbb{Z}$, the operator $Q_j(y)$ can be 
written as follows:
\begin{align}
	\label{eq:q}
	Q_j(y) = R_j^{y_1} \cdots R_2^{y_{j-1}} R_1^{y_j}.
\end{align}
Furthermore, from the equalities \ref{eq:r_identity} and \ref{eq:r_pow} it follows that:
\begin{align}
	Q_j(y)^{2^m}
		&= R_j^{y_1 \cdot 2^m} \cdots R_2^{y_{j-1} \cdot 2^m} R_1^{y_j \cdot 2^m} \nonumber \\
		&= R_{j - m}^{y_1} \cdots R_{2 - m}^{y_{j-1}} R_{1 - m}^{y_j} \nonumber \\
		&= Q_{j - m}(y).
\end{align}
The above equation implies that $Q_{j - m} = I$ if $j - m < 1$, therefore the $M_j(y, x)$ operator can be written as:
\begin{align}
	M_j(y, x) | 1 \rangle
		&= e^{2 \pi i \cdot 0.y_j \cdots y_1 \cdot x} | 1 \rangle \nonumber \\
		&= e^{2 \pi i \cdot 0.y_j \cdots y_1 \cdot (x_1 \cdot 2^{0} + x_2 \cdot 2^{1} + \cdots + x_k \cdot 2^{k - 1})} | 1 \rangle \nonumber \\
		&= Q_j(y)^{x_k \cdot 2^{k-1}} \cdots Q_j(y)^{x_2 \cdot 2^1} Q_j(y)^{x_1 \cdot 2^0} | 1 \rangle \nonumber \\
		&= Q_{j - k + 1}(y)^{x_k} \cdots Q_{j - 1}(y)^{x_2} Q_j(y)^{x_1} | 1 \rangle \nonumber \\
		&= Q_1(y)^{x_j} \cdots Q_{j-1}(y)^{x_2} Q_j(y)^{x_1} | 1 \rangle 
	\label{eq:Mj1} \\
	M_j(y, x) | 0 \rangle &= Q_1(y)^{x_j} \cdots Q_{j-1}(y)^{x_2} Q_j(y)^{x_1} | 0 \rangle = | 0 \rangle. 
	\label{eq:Mj2}
\end{align}

\begin{figure}[h!]
	\begin{center}
		\includegraphics[scale=0.3]{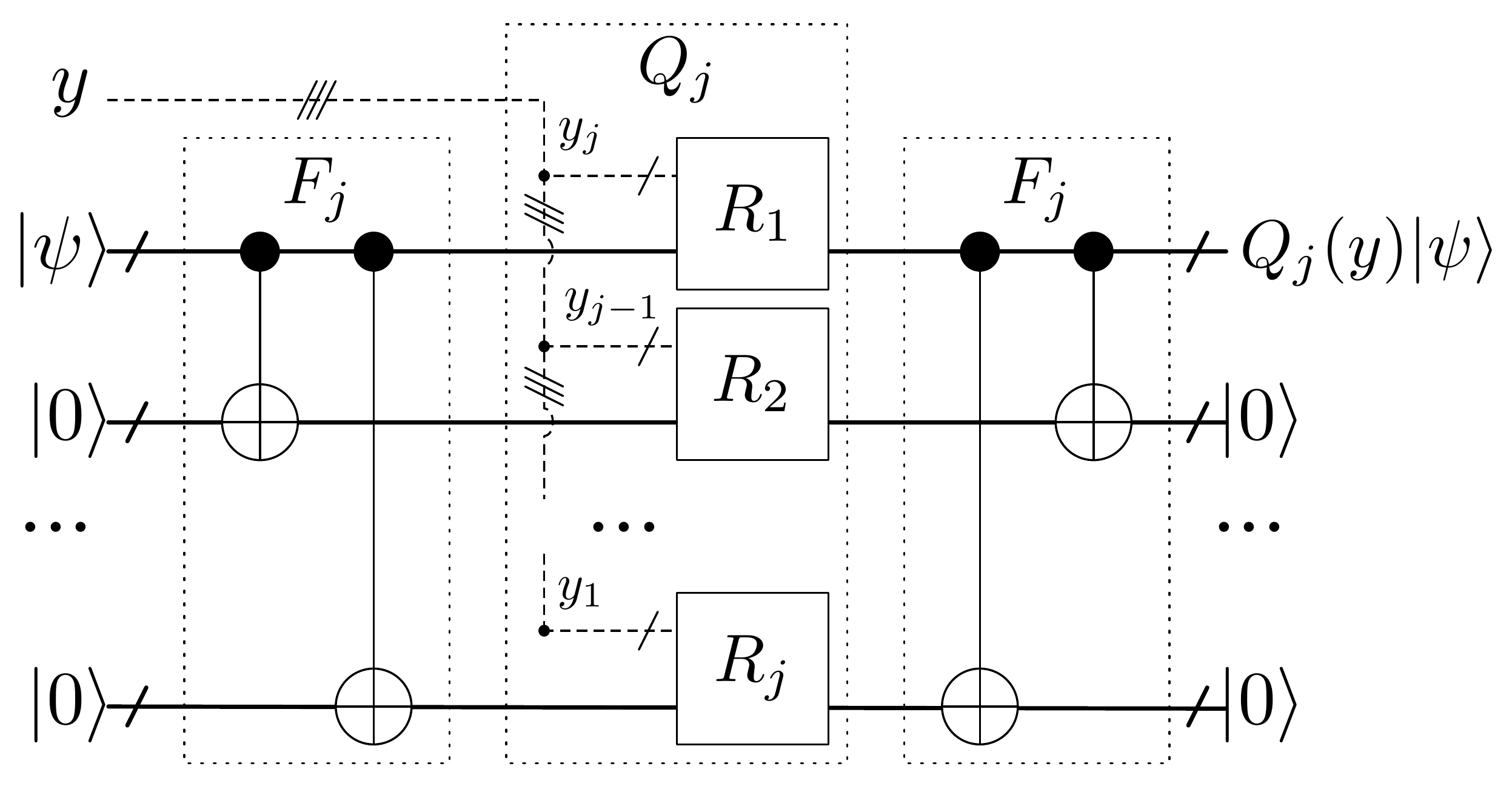}
	\end{center}
	\fcaption{\label{fig:q}The parallel version of the operator $Q_j(y)$. The $F_j$ blocks can be applied in $O(\log j)$ steps \cite{Moore:2001}. $|\psi\rangle$ is an arbitary 1 qubit state. The dashed lines represent classical bits and the continious lines qubits A single line crossing a wire represents a single bit/qubit and three lines crossing a wire represent multiple bits/qubits.}
\end{figure}

The decomposition of $M_j(y, x)$ into $Q_j(y)$ (Eq. \ref{eq:Mj1} and \ref{eq:Mj2}) operators and $Q_j(y)$ into
$R_j$ operators (Eq. \ref{eq:q}) will be used to construct a parallel quantum circuit for $M(y, x)$.
Note that the descriptions of $M(y, x)$, $M_j(y, x)$, and $Q_j(y)$ contain the arguments $x$ and $y$.
This is undesired for practical implementations of a circuit, since a circuit
cannot in general change depending on the input. In the design below,
this problem is resolved by using the bits of the arguments as controls
for quantum gates, \emph{i.e.} the value of classical bits is used to determine
if a particular quantum gate should be applied or not.
First, the parallel hybrid circuit for $Q_j(y)$ is constructed.
Since $R_j^0 = I$ and $R_j^1 = R_j$,
the effect of a input bit $y_m$ in Eq. \ref{eq:q}, where $m \in \{ 1, 2, \cdots, j \}$, is to control
the application of the gate $R_{j + 1 - m}$.
Thus the quantum circuit of $Q_j(y)$ can be
constructed using only single qubit $R_j$ gates controlled by classical bits $y_m$. 
All of the $R_j$ gates in $Q_j(y)$ can be applied in parallel using auxiliary qubits and
the $F_j$ gate \cite{Moore:2001}.
Thus the parallel hybrid circuit $F_j Q_j F_j$ of $Q_j(y)$ can be constructed as shown in Figure \ref{fig:q}.

\begin{figure}[h!]
	\begin{center}
		\includegraphics[scale=0.38]{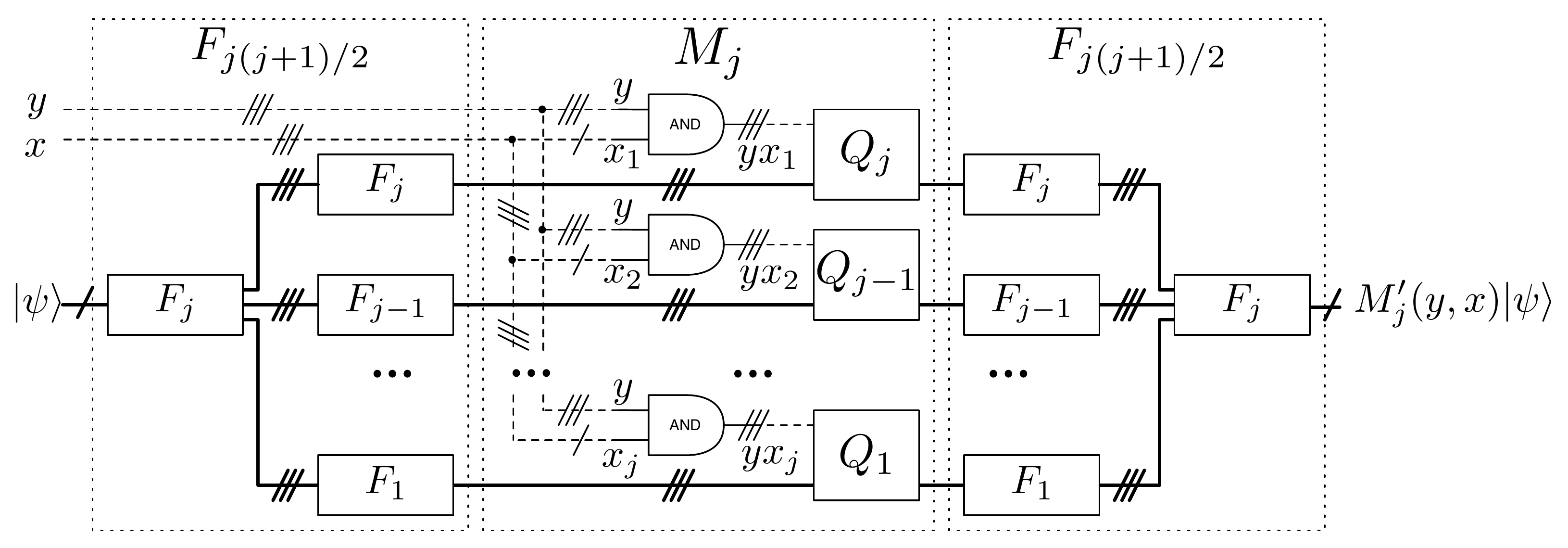}
	\end{center}
	\fcaption{\label{fig:Mj}The parallel hybrid circuit of the $M_j(y, x)$ operator.}
\end{figure}

\begin{figure}[h!]
	\begin{center}
		\includegraphics[scale=0.38]{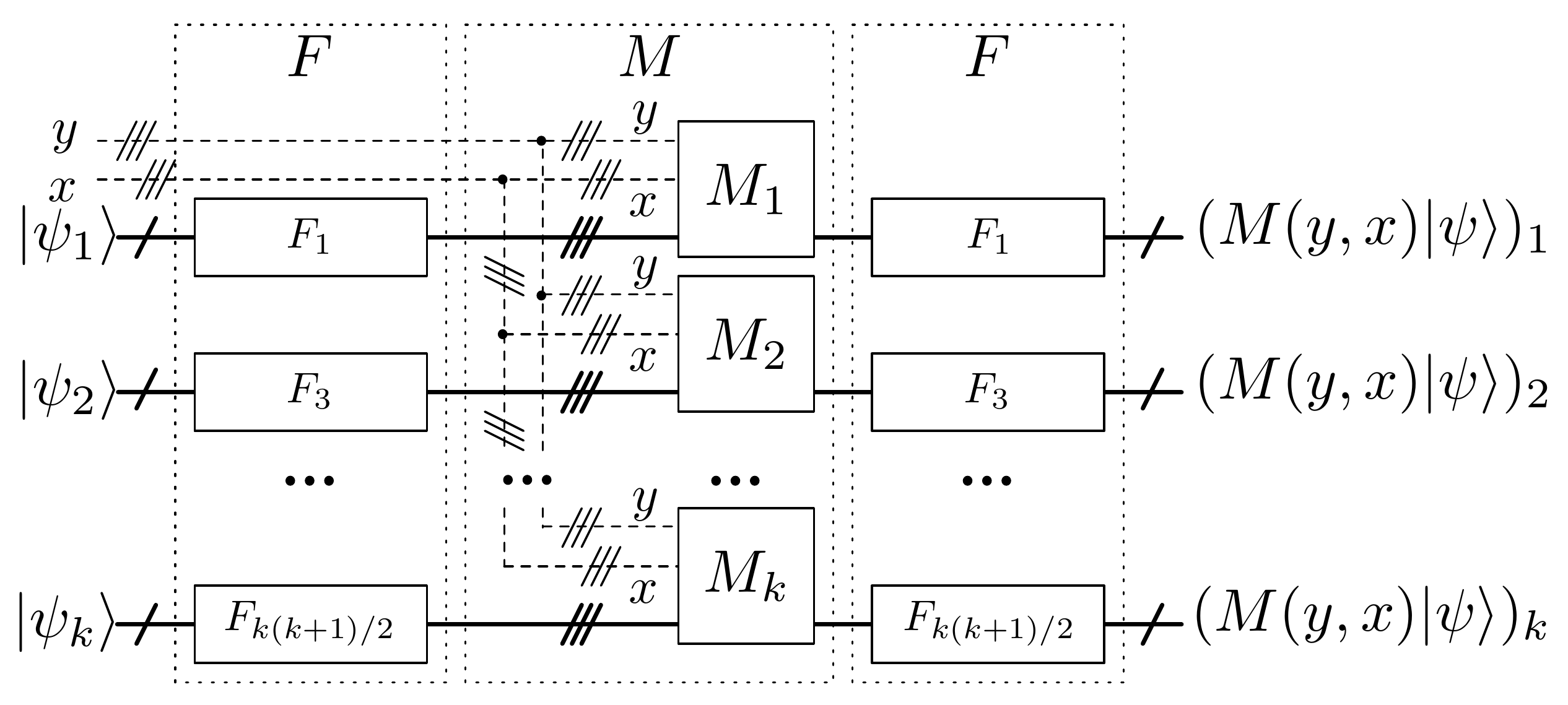}
	\end{center}
	\fcaption{\label{fig:M}The parallel hybrid circuit of the $M(y, x)$ operator.}
\end{figure}

Since $ Q_j(y)^0 = Q_j(0) = I$ and $x_i$ is a binary value, $M_j$ can
be written as $M_j(y, x) = Q_1(y \cdot x_j) \cdots Q_{j-1}(y \cdot
x_2) Q_j(y \cdot x_1)$.  The values of both $y$ and $x$ are classical
bit-strings, hence the operation $y \cdot x_i$ can be performed
classically in one parallel computational step using an AND operator
between $x_i$ and every bit of $y$. Since $M_j(y, x)$ can be
decomposed into diagonal operators $Q_j(y)$, there exists a parallel
hybrid circuit where all the $O_j(y)$ operators are applied
simultaneously \cite{Moore:2001}.  In this circuit's construction the
parallel hybrid circuit $O_j$ is used as seen in Figure \ref{fig:Mj}.
Since $M(y, x)$ is a tensor product of
the operators $M_j(y, x)$, where $j \in \{ 1, \cdots k \}$, the circuit of
$M(y, x)$ can be created by simply applying an appropriate
$M_j$ sub-circuit to each of the input qubits as shown in Figure \ref{fig:M}.
The circuit $FMF$ in the aforementioned figure corresponds to the operator $M(y, x)$
and together with the $QFT$ comprises the quantum MAC circuit.

\section{Analysis of the Circuit}
\label{sec:Analysis}

\begin{figure}
	\begin{center}
		\includegraphics[scale=0.4]{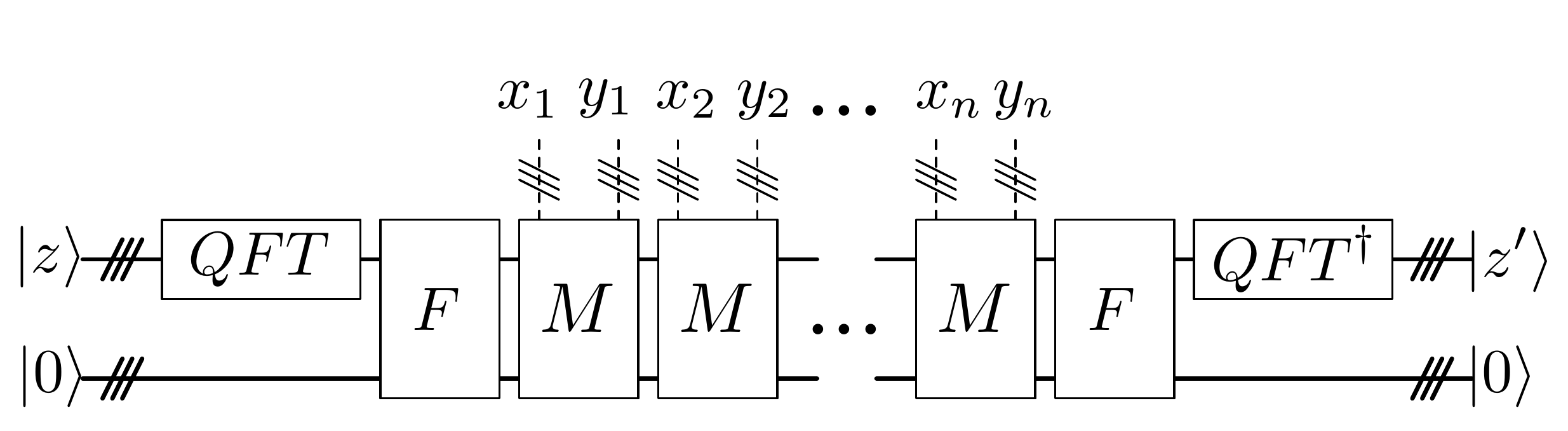}
	\end{center}
	\fcaption{\label{fig:QFTMAC} The hybrid quantum circuit computing the $MAC$ operation $n$ times in a sequence with multiplicand pairs $(x_1, y_1), \cdots, (x_n, y_n)$, where $z, x_i, y_i \in \mathbb{Z}$. Here $z' = z + \sum_{i=1}^{n} x_i \cdot y_i$.}
\end{figure}

\noindent
The main result of this work concerns the depth of the hybrid MAC
circuit in the case of sequential application. When the circuit $FMF$
in Figure \ref{fig:M} is applied in repeatedly, then the only $F$
gates having a non-trivial effect will be at the beginning and the end
of the computation. This is due to the fact that $FF = FF^\dagger = I$
and thus $(FMF)(FMF) = FMMF$. Combining the circuit in Figure
\ref{fig:M} with the $QFT$ and using it to perform the
multiply-addition operation of $n$ integers results in the hybrid circuit
depicted in Figure \ref{fig:QFTMAC}. As can be seen form the figure,
the overall depth will depend on the depth of $M$, which according
to the next lemma is constant.

\vspace*{12pt}
\begin{lemma}
	\label{lem:depth}
	The depth of the hybrid circuit $M$ is 2.
\end{lemma}
\vspace*{12pt}
\proof{
	It can be seen from figure \ref{fig:M} that the depth of the $M$
	circuit has to be equal to the maximum depth of any $M_j$
	sub-circuits, where $j \in \{ 1, \cdots, k \}$.  It is apparent that
	by substituting the $Q_j$ circuits in $M_j$, shown in Figure
	\ref{fig:Mj}, with the one described in Figure \ref{fig:q}, a
	circuit with one layer of classical $AND$ gates and one layer of
	single qubit $R_m$ gates can be constructed. Thus the combined depth of any $M_j$ and
	hence $M$, circuit is $2$
}

When determining the depth of a circuit, gates of variable size, such
as the $F$ gate have to be decomposed into one- and two-qubit quantum
gates.  An $F_m$ gate can be written as an $O(\log m)$ depth circuit
consisting of only $CNOT$ gates, where $m$ is the number of qubits
$F_m$ acts upon.  From Figure \ref{fig:M} it can be seen that the
number of qubits $F$ acts upon, is equal to the number of qubits $M$
acts upon. This in turn is equal to the number of quantum gates in $M$
since by Lemma \ref{lem:depth} there is only one layer
of quantum gates.  Thus the depth of the circuit in
\ref{fig:QFTMAC} it is given by the number of gates in $M$.

\vspace*{12pt}
\begin{lemma}
	\label{lem:size}
	The size of the hybrid circuit $M$ is $O(k^3)$.
\end{lemma}
\vspace*{12pt}
\proof{
	Let $size(C)$ be the size of a quantum circuit $C$, \emph{i.e.} the
	number of one- and two-qubit quantum gates in the decomposition of
	$C$.  Every $M_j$ sub-circuit in $M$ corresponds to one $M_j(y, x)$
	operator in the definition of $M(y,x)$.  Furthermore, every $Q_l$
	sub-circuit in $M_j$ corresponds to a $Q_l(y)$ operator in the
	definition of $M_j(y,x)$ (Eq. \ref{eq:Mj1} and \ref{eq:Mj2}) and each $R_m$ gate in
	$Q_l$ corresponds to a $R_m$ operator in the definition of $Q_l(y)$
	(Eq. \ref{eq:q}).  It can be seen from Eq. \ref{eq:q} that
	$size(Q_l) = l$ and the size of the circuit $M$ is therefore
	\begin{align}
		size(M) &= \sum_{j=1}^{k} size(M_j) = \sum_{j=1}^{k} \sum_{l=1}^{j} size(Q_l) = \sum_{j=1}^k\sum_{l=1}^j l = \sum_{j=1}^k \frac{j(j-1)}{2} = O(k^3).
	\end{align}
}

Now the overall depth of a hybrid circuit performing $n$ MAC
operations on $k$ bit integers can be estimated.  This is done for
both the exact and approximate case.

\vspace*{12pt}
\begin{theorem}
	There exists a hybrid quantum circuit with depth in $O(n + k)$ which performs $n$ multiply additions of $k$ bit integers exactly using $O(k^3)$ qubits.
\end{theorem}
\vspace*{12pt}
\proof{
  	The quantum circuit used to perform $n$ multiply additions is shown
  	in Figure \ref{fig:QFTMAC}. The number of qubits used by the circuit
  	is equal to the number of qubits the $M$ operator acts upon, which
  	by Lemma \ref{lem:size} is $O(k^3)$.
	
  	The $QFT$ and $QFT^\dagger$ of $k$ qubits can be applied in $O(k)$
  	depth \cite{Cleve:2000}. The fan-out operations $F$ can be
  	constructed using a tree-like structure so that the depth of that
  	circuit is logarithmic in the number of qubits to which they are applied\cite{Moore:2001},
  	\emph{i.e.} $O(\log k^3) = O(\log k)$. The $M$
  	circuit can be performed in exactly two steps as proven in Lemma
  	\ref{lem:depth} and it is applied it exactly $n$ times. Thus the overall
  	depth, \emph{i.e.} the number of parallel steps required for the
  	application of the circuit, is $n \cdot O(1) + 2 \cdot O(\log k) + 2
	\cdot O(k) = O(n + k)$
}

In practice it is unlikely that any quantum gates, or indeed,
classical logic gates, could be implemented perfectly. That is, there
will always be a small probability of the implemented gate failing,
resulting in a wrong answer. However, it is sufficient to obtain the
correct result with high enough probability.  When an exact result is
not required, the depth of a hybrid circuit computing multiple MAC
operations can be even smaller.  A
unitary operator is approximated with precision $\epsilon$ if for any
pure input quantum state the Euclidian distance between the desired
unitary $U$ and the implemented unitary $V$ is at most $\epsilon$.

\begin{theorem}
\vspace*{12pt}
	There exists a hybrid quantum circuit with depth $O(n + \log k + \log\log 1/\epsilon)$
	which performs $n$ multiply additions of $k$ bit integers with precision $\epsilon$ using $O(k^3 + k \cdot \log (k/\epsilon))$ qubits.
\end{theorem}
\vspace*{12pt}
\proof{
	To obtain a better depth than in the exact case a slightly modified
	version of the circuit in Figure \ref{fig:QFTMAC} is used in the approximate case.
	The initial $QFT$ can be replaced with a single layer of Hadamard gates applied to
	the $k$ qubit state $| 0 \rangle$. Note that this is equivalent to 
	applying $QFT$ to $| 0 \rangle$. Next the $M$ circuit is used to add $z \cdot 1$ to $QFT | 0 \rangle$,
	which gives us the state $QFT| 0 + z \cdot 1\rangle = QFT| z \rangle$.
	This is the same result as would be obtained by the exact circuit, but can be done in
	constant depth.
	
	An approximate version of $QFT^\dagger$, introduced in
        \cite{Cleve:2000}, can be used as the final step.  This
        $QFT^\dagger$ has depth $O(\log k + \log\log 1/\epsilon)$ and
        size $O(k \cdot \log (k/\epsilon))$ with precision $\epsilon$.
        The depth of the fan-out and $M$ operations are discussed above. The
        $M$ is applied exactly $n+1$ times. 
        Thus the overall depth is $(n + 1) \cdot O(1) + O(1) +
        2 \cdot O(\log k) + O(\log k + \log\log 1/\epsilon) = O(n +
        \log k + \log\log 1/\epsilon)$.
	
	The number of qubits used by the circuit is equal to the
        maximum number of qubits the $M$ operator acts upon, which
        by Lemma \ref{lem:size} is $O(k^3)$, and the number
        of qubits $QFT^\dagger$ acts upon. Thus the total number of
        qubits acted upon is $O(k^3 + k \cdot \log (k/\epsilon))$
}

\section{Results and Discussion}
\label{sec:Results}

\noindent
The depth of the proposed circuit for adding $n$ integers of $k$ bits
is $O(n + k)$ for the exact circuit and $O(n + \log k)$ for the
approximate circuit. The classical implementations of MAC are limited
by the depth complexity of the multiplication operations. This is
true even for the lowest depth multiplication circuits such as Wallace
\cite{Wallace:1964} and Dadda \cite{Dadda:1965} multipliers which are
used in most CPU architectures and have a depth of $O(\log k)$.  Thus
the sequential application of $n$ classical MACs requires at least
$O(n \log k)$ parallel steps.  It is unlikely that gate delays in
classical and quantum circuits will be the same. Indeed, they vary for
different classical circuits. However, in this analysis of the
different circuits the simple counting of the number of gates is
used. It is worth nothing that the advantage in depth gained by using
a QMAC increases with the number of sequential applications and the
size of integers used. Thus independent of the gate delays, there will be for
every integer size an $n$ such that performing at least $n$ MAC
operations has less depth when using the hybrid QMAC circuit than a
classical one.

The small depth of the QMAC is a consequence of using the QFT, a
highly entangled quantum state and classical fan-out, that is, copying
of bits.  First, since the MAC operation is performed on the QFT
state, only diagonal gates are necessary.  This makes it possible to
entangle the quantum register with auxiliary qubits in a way that
allows the simultaneous application of every single-qubit quantum
gate.  Second, the states of a bit can be copied by using multiple
output wires to more than two registers for the next computational
step.  Thus the information propagates in one step to all the
quantum gates controlled by these bits.  This can be interpreted as
influencing the state of an unbounded number of qubits with just one
fan-out operation.

The hybrid QMAC circuit implements a very specific computational
pattern, the MAC.  This makes it suitable for use as an execution unit
in a hybrid CPU or even a separate accelerator device such as a FPGA
or GPU in future computers. Moreover, the fact that implementing this
circuit does not require a full quantum computer makes it more likely
to be realisable in the near future. The small depth of the circuit
contributes to the ease of implementation, since the time needed to keep the
quantum state coherent depends on the circuit depth. A further consequence of the hybrid
nature of the circuit is that the number of qubits and two-qubit
gates used is relatively small. Instead of using only quantum
registers, two of the three registers in QMAC circuit are
classical. Using classically controlled single qubit gates instead of
fully quantum controlled gates limits the number of two-qubit gates
used. However, the entangled state used
requires $O(k^3)$ auxiliary qubits and two-qubit gates.

Future work would be to consider how to adapt the hybrid QMAC circuit
floating to point operations, which are used in most of the
time-intensive computations. This would greatly increase the number of
problems which would benefit from quantum devices.  Another direction
would be to consider hybrid circuits for other arithmetic operations
for example division and look at how the different circuits can
be combined together. The QMAC introduced in this paper has a lower
depth than a classical MAC only if it is applied in a sequence.  Hence
combining different quantum arithmetic operators could result in an
improved depth compared with classical circuits.

\nonumsection{References}
\bibliographystyle{unsrt}
\bibliography{bibliography}

\end{document}